**Resonant tunnelling and negative differential conductance in graphene transistors**

L. Britnell[1], R. V. Gorbachev[2], A. K. Geim[1,2], L. A. Ponomarenko[1], A. Mishchenko[1],

M. T. Greenaway[3], T. M. Fromhold[3], K. S. Novoselov[1] & L. Eaves[1,3]

[1]School of Physics & Astronomy, University of Manchester, Manchester M13 9PL, UK

[2]Manchester Centre for Mesoscience & Nanotechnology, University of Manchester, Manchester M13 9PL, UK

[3]School of Physics & Astronomy, University of Nottingham, Nottingham NG7 2RD, UK

**The chemical stability of graphene and other free-standing two-dimensional crystals means that they can be stacked in different combinations to produce a new class of functional materials, "tailor-made" for device applications. Here we report resonant tunnelling of Dirac fermions through a boron nitride barrier, a few atomic layers thick, sandwiched between two graphene electrodes. The resonant peak in the device characteristics occurs when the electronic spectra of the two electrodes are aligned. The resulting negative differential conductance persists up to room temperature and is gate voltage-tuneable due to graphene's unique Dirac-like spectrum. Whereas conventional resonant tunnelling devices comprising a quantum well sandwiched between two tunnel barriers are tens of nanometres thick, the tunnelling carriers in our devices cross only a few atomic layers, offering the prospect of ultra-fast transit times. This feature, combined with the multi-valued form of the device characteristics, has potential for applications in high-frequency and logic devices.**

# **Introduction**

Multilayer stacks of graphene and other stable, atomically thin, two-dimensional materials such as boron nitride, the metallic dichalcogenides and layered oxides[1] offer the prospect of creating a new class of heterostructure materials. Devices based on these materials have been investigated in a number of recent experimental and theoretical papers[2-28]. By fine-tuning the composition of the stacks, it may be possible to create materials with novel electronic and optical properties, which outperform conventional semiconductors. In this article, we use this technology to fabricate and study a resonant tunnelling device in which carriers tunnel through a thin boron nitride barrier layer sandwiched between two graphene electrodes across which a bias voltage is applied. We mount the layers on the oxidised surface of a doped silicon substrate, which we use as a gate electrode, The additional tuning provided by the gate electrode allows us to probe in detail the physics of resonant tunnelling of Dirac-fermions and to exploit graphene's unique electronic spectrum, whereby the carrier type (electron or hole) and sheet density can be controlled by the electrostatic field effect[12,23].

We observe a strong resonant peak in the current-voltage characteristics of the device over a wide range of gate voltages with peak-to-valley ratios of up to 4-to-1. The peak and resulting negative differential conductance (NDC) persist up to room temperature. A model for the carrier tunnelling based on a transfer Hamiltonian approach provides a good fit to the measured device characteristics.

# Results

**Device structure**

The structure of our devices is shown schematically in Figure 1(a). A thick layer of hBN is placed on top of an oxidised Si substrate and acts as an atomically-flat substrate[2] on which the active part of the device is mounted. This consists of two graphene electrodes on either side of an atomically thin hBN tunnel barrier. A tunnel current is generated when we apply a bias voltage, $V_b$, between the bottom and top graphene electrodes; the gate voltage, $V_g$, is applied between the doped silicon substrate and bottom electrode. This arrangement allows us to align the Dirac points of the two graphene electrodes whilst maintaining control of their chemical potentials, as shown schematically in the inset of Figure 1(b). Due to the low density of states close to the Dirac point, the bottom graphene electrode only partially screens the charge that is induced on the Si-gate electrode by $V_g$; hence the sheet carrier density in the top electrode is also influenced by $V_g$. This behaviour is related to the so-called quantum capacitance[29], which is a strong effect in graphene, unlike the case of conventional semiconductors. A relatively modest $V_g$ is sufficient to induce a large change in the chemical potential, $\mu_B$, of the bottom graphene layer, thus providing additional control of the effective barrier height and transmission coefficient[12,23].

**Low temperature measurements**

The main body of Figure 1(b) shows the measured gate-dependent resonant tunnelling characteristics for one of our devices, device A, in which the hBN barrier is four atomic layers thick. Unless stated otherwise, the following discussion of the physics and modelling is focussed on measured characteristics of this device. The most notable feature of the data is the presence of a pronounced peak in the current, whose amplitude and voltage position are sensitive to $V_g$. Beyond the peak, there is an extended region of NDC. This high-bias regime was not explored in earlier work on graphene-based tunnelling transistors[12]. The $I(V_b)$ curve is approximately symmetric at $V_g \approx -20$ V (black curve), but becomes increasingly asymmetric as $V_g$ is tuned away from this voltage. Whereas the peak-to-valley ratio (PVR) is relatively small for the symmetric $I(V_b)$ curve, it is significantly enhanced by tuning $V_g$, reaching a value of $\approx 2$ at $V_g = -55$ V. Note also that the $V_g$-induced increase in the peak height and PVR for one polarity of $V_b$ is accompanied by a corresponding decrease for the opposite polarity.

**Theoretical model**

To make a direct comparison with the experimental data, Figure 1(c) shows theoretical simulations based on the Bardeen transfer Hamiltonian approach, which includes the effects of residual doping in the graphene electrodes on the device characteristics. Before discussing this figure in detail, let us first consider an idealised case in which the graphene electrodes are undoped and their crystal lattices are aligned. If $V_g = V_b = 0$, the Fermi energy μ in each layer is at the Dirac point and the two Dirac points are at the same energy, as shown in inset (a) of Figure 2. If $V_b > 0$, electrons accumulate in the negatively-biased electrode with an equal number of holes in the positive electrode (inset (b) of Figure 2). This charging of the two layers generates an electric field which misaligns the two Dirac points. Therefore, in order for a carrier to tunnel from one electrode to the other, its in-plane wavevector *k* must change. Such a change is forbidden unless the tunnelling event is accompanied by a scattering process in which the conservation of *k* is relaxed. Figure 2, inset (c), shows the effect of a positive $V_g$ on the Fermi energies in the two graphene electrodes when $V_b = 0$ so that the two chemical potentials are aligned. Keeping $V_g$ fixed, we can adjust $V_b$ to bring the Dirac points of the two electrodes into alignment, see Figure 2 inset (d), thus allowing all carriers whose energies are between the now distinct chemical potentials of the two electrodes to tunnel resonantly.

In real devices, elastic scattering due to disorder or interaction effects leads to relaxation of the *k* conservation condition, so that the resonant feature which appears in the current-voltage curves has a line-width and position that depend on the disorder potential or the nature of the interaction. This type of behaviour has been investigated in conventional III-V heterostructure devices in which carriers tunnel resonantly through a barrier sandwiched between two confining quantum wells, so-called "2D-to-2D tunnelling"[30,31].

For a tunnelling electron, a disorder-induced change, *q*, in its in-plane wavevector can be described in terms of a scattering potential, $V_S(q)$. If $V_S(q)$ is weakly dependent on *q*, the tunnel current increases monotonically with $V_b$, so no peak is observed. However, a resonance can appear if $V_S(q)$ has a bell-shaped dependence, falling towards zero when *q* exceeds a certain value $q_c$, where $q_c^{-1}$ is the lower limit of the modulation length of the disorder-induced scattering potential in real space. To model our data, we use a short-range scattering potential, $V_S(q) \propto 1/(q_c^2 + q^2)$, where $q_c^{-1} = 12$ nm. Such a $V_S(q)$ can arise from potential variations associated with electron-hole puddles[32], short range disorder[33,34], the Moiré pattern at graphene-hBN interfaces[10,35,36] (there are three such interfaces in our device) and/or acoustic phonon scattering. The simulations shown in Figure 2 are qualitatively similar to those observed for our devices. The region of NDC corresponds approximately to $V_b \sim \hbar v_F q_c/e$, the situation where the typical in-plane wavevector change, *q*, approaches $q_c$; $v_F$ is the Fermi velocity in graphene.

**Theoretical model of the measured device characteristics**

We now focus on the quantitative features in the measured $I(V_b)$ curves which are not reproduced in the simulations using undoped graphene layers. Note, firstly, that the measured differential conductance at $V_b = 0$ is finite at all values of $V_g$, see Figure 1(b), whereas in Figure 2 it tends to zero at $V_b = 0$ for the (black) curve with $V_g = 0$. Secondly, the measured $I(V_b)$ curves are symmetric at $V_g = -20$ V, whereas for the model calculation in Figure 2 the curve is symmetric at $V_g = 0$. Our model successfully reproduces these features when we take into account the effect of unintentional, residual doping of the graphene layers. As can be seen by comparing Figure 1(b) and (c), good quantitative agreement between experiment and theory for device A is obtained when the residual doping has the values stated in the figure caption. All the essential features of the measured $I(V_b)$ characteristics are then reproduced: the $V_g$-dependence of the linear conductance around zero bias due to the finite density of states – see inset (i) in Figure 1(c); the $V_g$-induced shift of the peak in current and the NDC region beyond; the way in which the application of $V_g$ enhances the peak in $I(V_b)$ and associated NDC in one bias direction, whilst decreasing it in the other bias direction.

The strong peak in $I(V_b)$ arises from the resonant alignment of the energies of Dirac cones in the two electrodes and the large number of empty states in the top electrode which are available to electrons tunnelling from the bottom emitter layer, as illustrated in inset (ii) of Fig. 1(c). The lower PVR in the measured $I(V_b)$ compared to those in our calculations may arise in part from small leakage currents between the graphene electrodes involving other tunnelling mechanisms or a more complex $V_S(q)$ with less sharp cut-off than assumed in our model. Our analysis also explains the origin of the shoulder-like feature beyond the resonant peak in reverse bias: the shoulder occurs when $V_b$ and $V_g$ correspond to the alignment of the chemical potential in the top graphene layer with the Dirac point of the bottom layer. When this condition is satisfied, the carriers close to the Fermi energy in one electrode tunnel into a very low density of states in the other, so that the measured current is insensitive to small changes of $V_b$, thus yielding a differential conductance close to zero.

**Resonant tunnelling and NDC at room temperature**

Figure 3, showing data for another transistor, device B, illustrates the reproducibility of the *I-V* characteristics for different devices and the persistence of the resonant tunnelling effect up to room temperature, a key property for future device applications. In this device, a PVR of up to ≈4 is obtained at low *T* (see inset of Fig. 3(a)) and the resonant peak and negative differential conductance remain clearly defined at room temperature (Fig. 3(b)). The dependence of the PVR on $V_g$ (inset of Fig. 3(a)) is qualitatively similar to device A and closely reproduces that obtained from the model (bottom right inset of Fig. 1(c)).

# Discussion

Several different mechanisms by which negative differential conductance can be generated in graphene-based devices have been modelled theoretically, e.g. refs. 3, 18 and 27. The observation of NDC in a graphene field effect tunnel transistor in which the current flows along the graphene layer was also reported recently. In this device, a tunnel barrier potential is formed in the plane by applying a gate voltage across a silicon nitride dielectric stripe deposited on a single graphene layer[37]. Our device, in which the tunnel current flows perpendicular to the plane of the component layers, compares very favourably with the planar device, particularly with regard to its strong NDC and high peak-to-valley ratio. It is also interesting to compare our device with resonant tunnelling in double barrier devices made from conventional semiconductor heterostructures[38,39]. In those devices, the quantum well confinement provided by the two potential barriers creates quasi-two-dimensional states through which charged carriers can tunnel when their energy is tuned to resonance by an applied voltage. In contrast, our devices consist of a single tunnel barrier, so the device speed is not limited by the dwell time of the carriers in a central quantum well[40,41]. Although the peak-to-valley ratio of our proof-of-concept devices is significantly lower than that achieved in state-of-the-art III-V double barrier resonant tunnelling diodes[42,43], there are good prospects of achieving higher peak-to-valley ratios. In addition, since the barrier transmission coefficient is exponentially sensitive to the number of atomic layers in the hBN barrier[13], it should be possible to achieve much higher peak resonant current densities in devices by using, for example, three-monolayer barriers.

We note the recent theoretical work by Feenstra *et al.*[15,26], which modelled carrier tunnelling in multilayer graphene-dielectric-graphene tunnel diodes. That work considered how crystal grain size and spatial misorientation of the lattices of the two graphene electrodes can influence resonant tunnelling. Microstructural analysis[2,12,13] of the graphene and boron nitride layers that make up our devices indicates that they have a high level of structural perfection, so grain size effects are unlikely to have a major influence on the behaviour of our devices. Supplementary note 1 discusses how we incorporate the effects of misorientation of the graphene lattices in our model of the device characteristics.

In summary, we have shown how to achieve gate-controlled resonant tunnelling with high peak-to-valley ratio and pronounced negative differential conductance in graphene-based multilayer, "vertical" transistors. These devices could be developed further for applications in logic circuits that exploit the multi-valued form of the current-voltage characteristics. The additional functionality provided by resonant tunnelling is not available in graphene "barristors"[23] in which the current flow arises from carriers that are thermally activated over a Schottky potential barrier. Furthermore, one of the key factors that limits the speed of conventional double barrier resonant tunnelling diodes, namely the dwell time of carriers in the central quantum well at resonance, is avoided in our proof-of-concept device architecture. This, combined with the high mobility of the carriers in the graphene electrodes and the atomically thin tunnel barrier, suggests potential applications in high-speed electronics.

## Methods

**Device fabrication**

The methods used for fabricating this type of the device are given in references [12] (and its supporting online material) and [13].

**Measurement**

The device characteristics were measured using conventional electrical and cryogenic techniques.

**Theoretical model**

The theoretical model used to fit our data is described in supplementary note 1.

**References**

1. Novoselov, K. S. *et al.* Two-dimensional atomic crystals. *PNAS* **102**, 10451-10453 (2005).

2. Dean, C. R. *et al.* Boron nitride substrates for high-quality graphene electronics. *Nature Nano.* **5**, 722-726 (2010).

3. Ferreira, G. J., Leuenberger, M. N., Loss, D. and Egues, J. C. Low-bias negative differential resistance in graphene nanoribbon superlattices. *Phys. Rev. B* **84**, 125453 (2011).

4. Karpan, V. M., Khomyakov, P. A., Giovanetti, G. *et al.*, Ni(111)□graphene□*h*-BN junctions as ideal spin injectors. *Phys. Rev. B* **84**, 153406 (2011).

5. Ponomarenko, L. A. *et al.* Tunable metal-insulator transition in double-layer graphene heterostructures. *Nature Phys.* **7**, 958-961 (2011).

6. Qiu, M. and Liew, K. M. Transport properties of a single layer armchair h-BNC heterostructure. *J. Appl. Phys.* **110**, 064319 (2011).

7. Sciambi, A. *et al.* Vertical field-effect transistor based on wave-function extension. *Phys. Rev. B* **84**, 085301 (2011).

8. Wang, H. *et al.* BN/graphene/BN transistors for RF applications. *IEEE Elec. Dev. Lett.* **32**, 1209-1211 (2011).

9. Xu, Y., Guo, Z. D., Chen, H. B., Yuan, Y. *et al.* In-plane and tunneling pressure sensors based on graphene/hexagonal boron nitride heterostructures. *Appl. Phys. Lett.* **99**, 133109 (2011).

10. Xue, J. M. *et al.* Scanning tunnelling microscopy and spectroscopy of ultra-flat graphene on hexagonal boron nitride. *Nature Mater.* **10**, 282–285 (2011).

11. Amet, F. *et al.* Tunneling spectroscopy of graphene-boron-nitride heterostructures. *Phys. Rev. B* **85**, 073405 (2012).

12. Britnell, L. *et al.* Field-effect tunneling transistor based on vertical graphene heterostructures. *Science* **335**, 947-950 (2012).

13. Britnell, L. *et al.* Electron tunneling through ultrathin boron nitride crystalline barriers. *Nano Letters* **12**, 1707-1710 (2012).

14. Dean, C. *et al.* Graphene based heterostructures. *Solid State Comm.* **152**, 1275-1282 (2012).

15. Feenstra, R. M., Jena, D. & Gu, G. Single-particle tunneling in doped graphene-insulator-graphene junctions. *J. Appl. Phys.* **111**, 043711 (2012).

16. Fiori, G., Betti, A., Bruzzone, S. and Iannaccone, G. Lateral graphene-hBCN heterostructures as a platform for fully two-dimensional transistors. *ACS Nano* **6**, 2642-2648 (2012).

17. Haigh, S. J. *et al.* Cross-sectional imaging of individual layers and buried interfaces of graphene-based heterostructures and superlattices. *Nature Mater.* **11**, 764-767 (2012).

18. Hung Nguyen, V., Mazzamuto, F., Bournel, A. & Dollfus, P. Resonant tunnelling diodes based on graphene/h-BN heterostructure. *J. Phys. D: Appl. Phys.* **45**, 325104 (2012).

19. Kumar, S. B., Seol, G., Guo J. Modeling of a vertical tunneling graphene heterojunction field-effect transistor. *Appl. Phys. Lett.* 101, 033503 (2012).

20. Levendorf, M. P. *et al.* Graphene and boron nitride lateral heterostructures for atomically thin circuitry. *Nature* **488**, 627-632 (2012).

21. Lopez-Bezanilla, A. and Roche, S. Embedded boron nitride domains in graphene nanoribbons for transport gap engineering. *Phys. Rev. B.* **86**, 165420 (2012).

22. Kim, S. and Tutuc, E. Magnetotransport and Coulomb drag in graphene double layers. *Solid State Commun.* **15**, 1283-1288 (2012).

23. Yang, H. *et al.* Graphene barristor, a triode device with a gate-controlled Schottky barrier. *Science* **336**, 1140-1142 (2012).

24. Georgiou, T. *et al.* Vertical field-effect transistor based on graphene-$WS_2$ heterostructures for flexible and transparent electronics. *Nature Nanotech.* **8,** 100-103 (2013).

25. Bruzzone, S., Fiori, G. and Iannaconne, G. Tunneling properties of vertical heterostructures of multilayer hexagonal boron nitride and graphene. *Preprint* arXiv:1212.4629v1 (2012).

26. Zhao, P., Feenstra, R. M., Gu, G. & Jena, D., SymFET: A proposed symmetric graphene tunneling field effect transistor. *IEEE Trans. Electron Devices* **60,** 951-957 (2013).

27. Nam Do, V., Hung Nguyen, V., Dollfus, P. and Bournel, A. Electronic transport and spin-polarization effects of relativisticlike particles in mesoscopic graphene structures. *J. Appl. Phys.* **104**, 063708 (2008).

28. Vasko, F. T., Resonant and nondissipative tunnelling in independently contacted graphene structures. *Phys. Rev. B,* **87,** 075424 (2013).

29. Luryi, S. Quantum capacitance devices. *Appl. Phys. Lett.* **52**, 501-503 (1988).

30. Eaves, L. *et al.* Electrical and spectroscopic studies of space-charge buildup, energy relaxation and magnetically enhanced bistability in resonant-tunneling structures. *Solid State Electronics* **32**, 1101-1108 (1989).

31. Eisenstein, J. P., Pfeiffer, L. N. & West, K. W. Coulomb barrier to tunneling between parallel 2-dimensional electron-systems. *Phys. Rev. Lett.* **69**, 3804-3807 (1992).

32. Martin, J. *et al.* Observation of electron-hole puddles in graphene using a scanning single-electron transistor. *Nature Phys.* **4**, 144-148 (2008).

33. Qiuzi Li, Hwang, E. H., Rossi, E. & Das Sarma, S. Theory of 2D transport in graphene for correlated disorder. *Phys. Rev. Lett.* **107**, 156601 (2011).

34. Yan, J., Fuhrer, M. S. Correlated charged impurity scattering in graphene. *Phys. Rev. Lett.* **107**, 206601 (2011).

35. Decker, R. *et al.* Local electronic properties of graphene on a BN substrate via scanning tunneling microscopy. *Nano Lett.* **11**, 2291-2295 (2011).

36. Yankowitz, M. *et al.* Emergence of superlattice Dirac points in graphene on hexagonal boron nitride. *Nature Phys.* **8**, 382-386 (2012).

37. Wu, Y., Farmer, D. B., Zhu, W., Han, S.-J. *et al.*, Three-terminal graphene negative differential resistance devices. *ACS Nano* **6**, 2610-2616 (2012).

38. Chang, L. L., Esaki, L. & Tsu, R. Resonant tunneling in semiconductor double barriers. *Appl. Phys. Lett.* **24**, 593-595 (1974).

39. For a review, see Mizuta, H. & Tomonori, T. *The Physics and Applications of Resonant Tunnelling Diodes* (Cambridge Univ. Press, Cambridge, 1995).

40. Suzuki, S., Asada, M., Teranishi, A., Sugiyama, H. & Yokoyama, H. Fundamental oscillation of resonant tunneling diodes above 1 THz at room temperature. *Appl. Phys. Lett.* **97**, 242102 (2010).

41. Feiginov, M., Sydlo, C., Cojocari, O. & Meissner P. Resonant-tunneling-diode oscillators operating at frequencies above 1.1 THz. *Appl. Phys. Lett.* **99**, 233506 (2011).

42. Day, D. J., Chung, Y., Webb, C., Eckstein, J. N. *et al.* Double quantum well resonant tunnel diodes. *Appl. Phys. Lett.* **77**, 1260-1261 (1990).

43. Tsai, H. H., Su, Y. K., Lin, H. H. *et al.*, P-N double quantum well resonant interband tunneling diode with peak-to-valley current ratio of 144 at room temperature. *IEEE Electron Device Lett.* **15**, 357-359 (1994).

**Acknowledgements**

This work was supported by the European Research Council, Engineering and Physical Research Council (UK), U.S. Office of Naval Research, U.S. Air Force Office of Scientific Research, European Commission FP7, and the Körber Foundation.

**Author contributions**

AKG, KSN and LE conceived and designed the experiments, which were performed by LB, RVG, LAP and AM. AKG, MTG, TMF, KSN and LE analysed the data; MTG, TMF and LE wrote the supplementary information and developed the model. MTG wrote code and ran the model. The Letter was written by AKG, TMF and LE with contributions from all authors.

**Competing financial interests**

The authors declare no competing financial interests.

**Figure Captions**

**Figure 1: Graphene-BN resonant tunnelling transistor.**

(a) Schematic diagram of the devices.

(b) Measured current-voltage characteristics of one of our devices (device A) at 6 K. The hBN barrier is 4 atomic layers thick, as determined by atomic force microscopy and optical contrast[13]; the active area for the flow of tunnel current is 0.3 µm$^2$. The $V_g$ values range, in 5 V steps, from +15 V (top red curve), through -20 V (black symmetric curve) to -55 V (bottom blue curve). The inset shows schematically the relative positions of the Fermi energies (chemical potentials) of the doped Si substrate gate electrode (represented by hatched lines) and of the two graphene layers at the peak of the $I(V_b)$ curve in forward bias with $V_g = +15$ V.

(c) Theoretical simulation of device A obtained by using the Bardeen model and including the effect of doping in both graphene electrodes. Parameters: $q_c^{-1} = 12$ nm; bottom layer is n-doped at $4.4 \times 10^{11}$ cm$^{-2}$ and the top graphene p-doped at $1.0 \times 10^{12}$ cm$^{-2}$. Since our top graphene layers are exposed to the environment, we expect them to have stronger residual (~$10^{12}$ cm$^{-2}$) doping than the bottom layers, as often observed in partially encapsulated double-layer graphene devices[5]. The top inset (i) shows the chemical potentials $\mu_T$ and $\mu_B$ in the top (T) and bottom (B) electrodes, respectively, for $V_b = 0$ and $V_g = -20$ V, which corresponds to the symmetric $I(V_b)$ shown in black; for inset (ii) $V_g = +15$ V and $V_b = 0.3$ V, which corresponds to the peak of the $I(V_b)$ curve. The lower inset shows the $V_g$-dependence of the PVR = $I_p/I_v$ obtained from our simulations, where $I_{p,v}$ are the currents at the peak and the valley (minimum) beyond.

**Figure 2: Idealized current-voltage characteristics of a resonant tunnel transistor.** Here, we assume that the chemical potentials of the two graphene electrodes are at the Dirac points when the device is unbiased. $T = 10$ K. $V_g = 0$ (black curve); $V_g = 10$ V (red curve). Insets (a) to (d) show the positions of the chemical potentials at the marked points on the $I(V_b)$ curves.

**Figure 3: Reproducibility of resonant tunnelling in double-layer graphene devices.** Device B exhibits $I(V_b)$ curves similar to those of device A. The hBN barrier is 5 atomic layers thick; the active area for the flow of tunnel current is 0.6 µm². (a) $T = 7$ K; the $V_g$ values range, in 5 V steps, from +20 V to -60 V. The inset shows the $V_g$-dependence of the PVR. (b) Room-temperature $I(V_b)$ characteristics for device B; $V_g$ ranges from +50 to -50 V in 5 V steps.

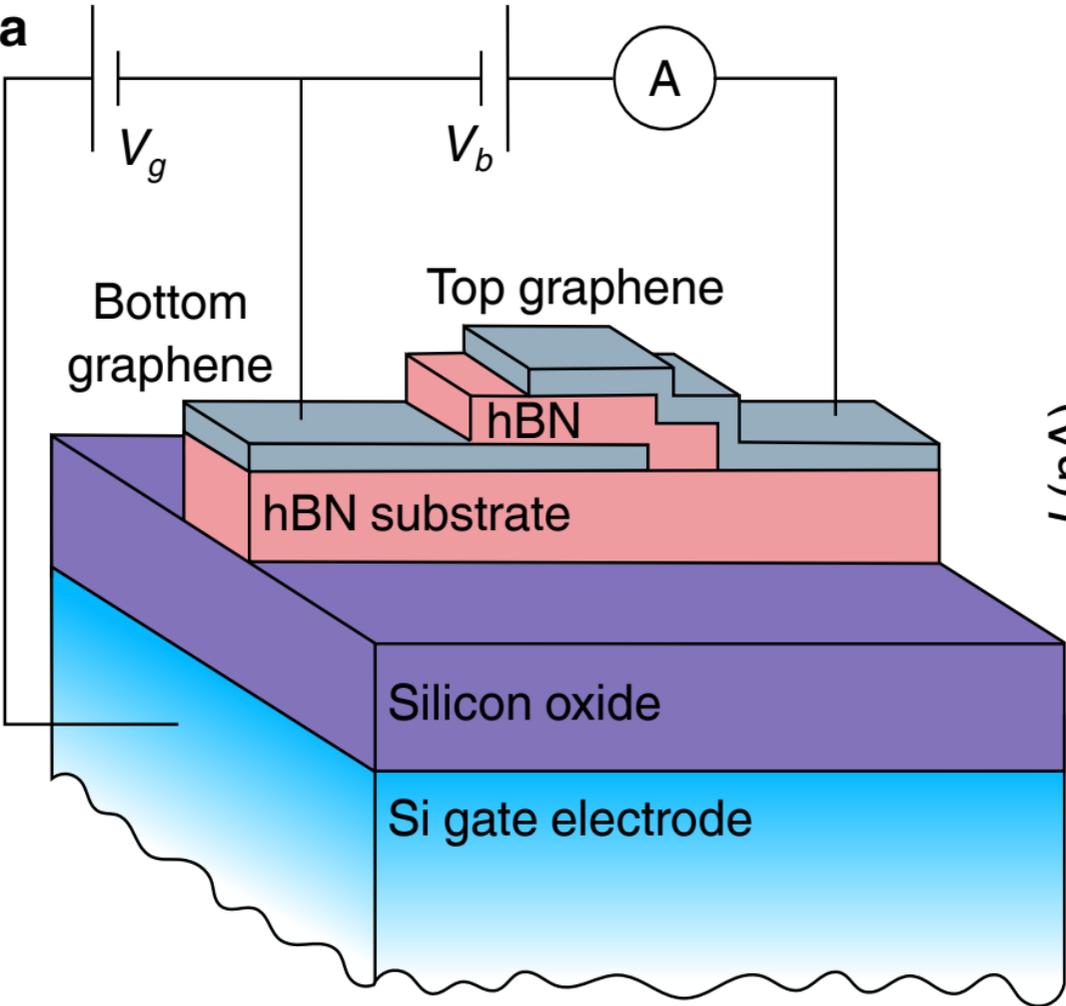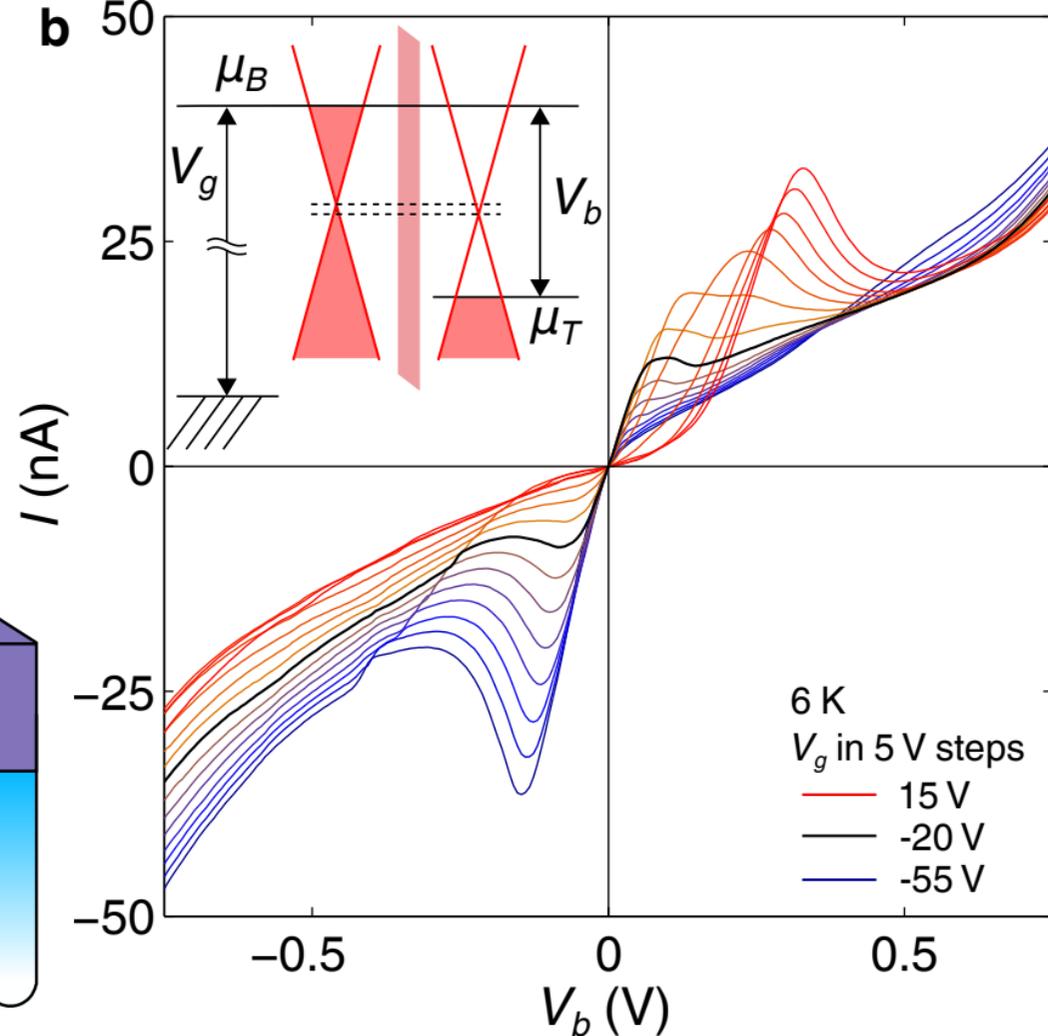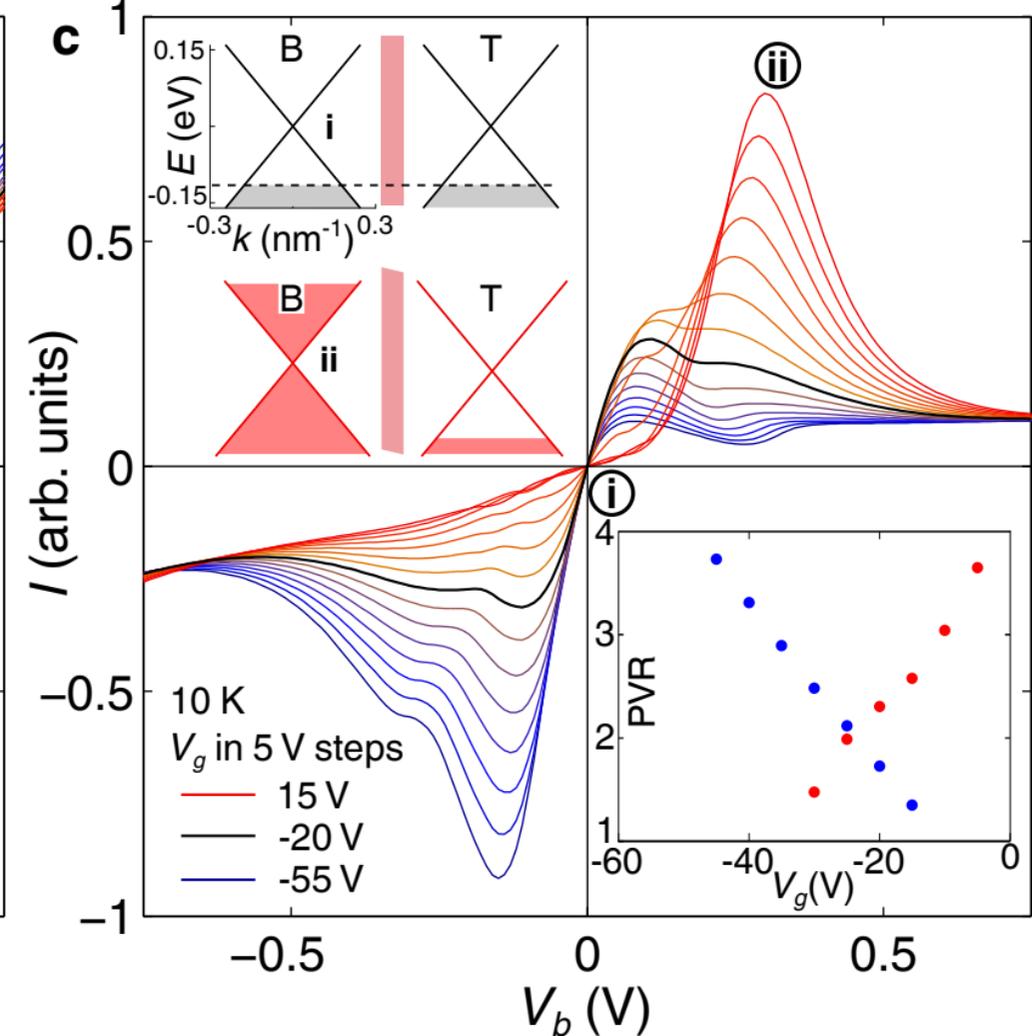

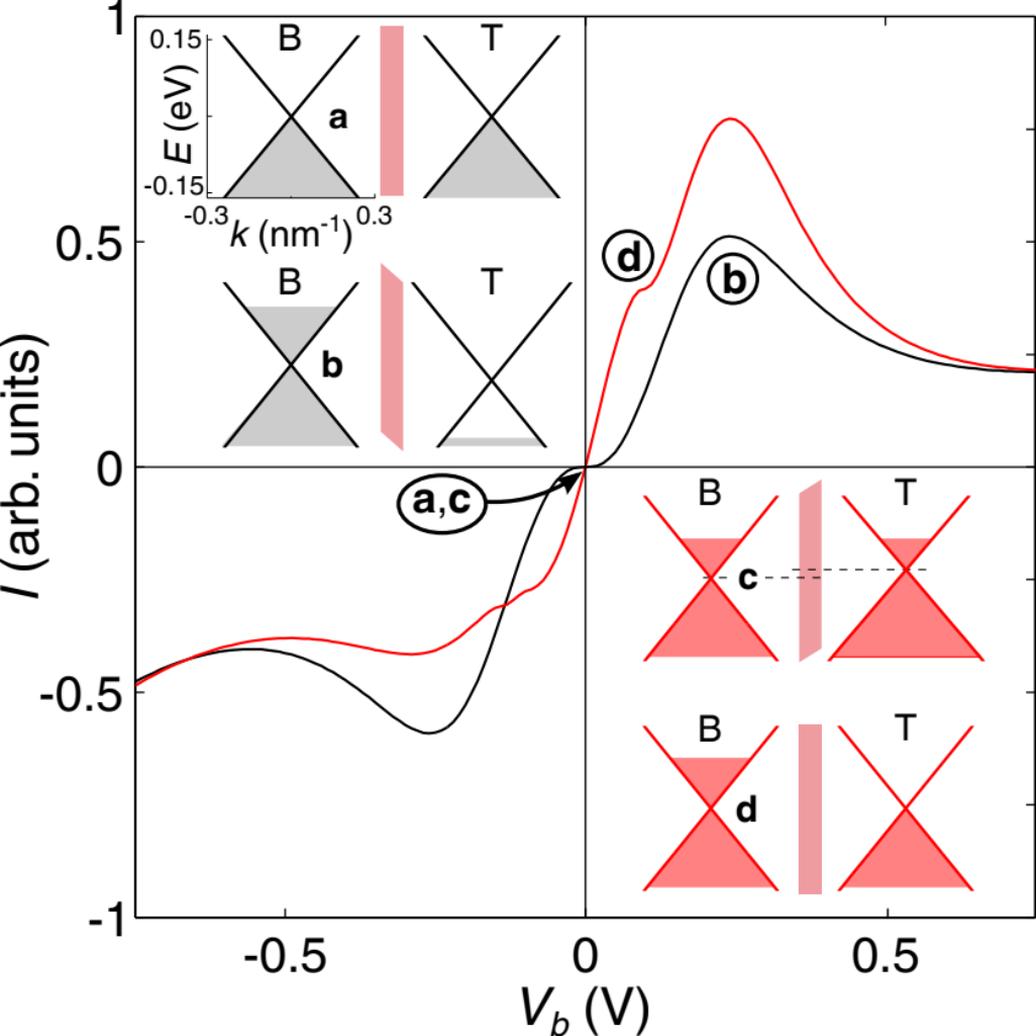

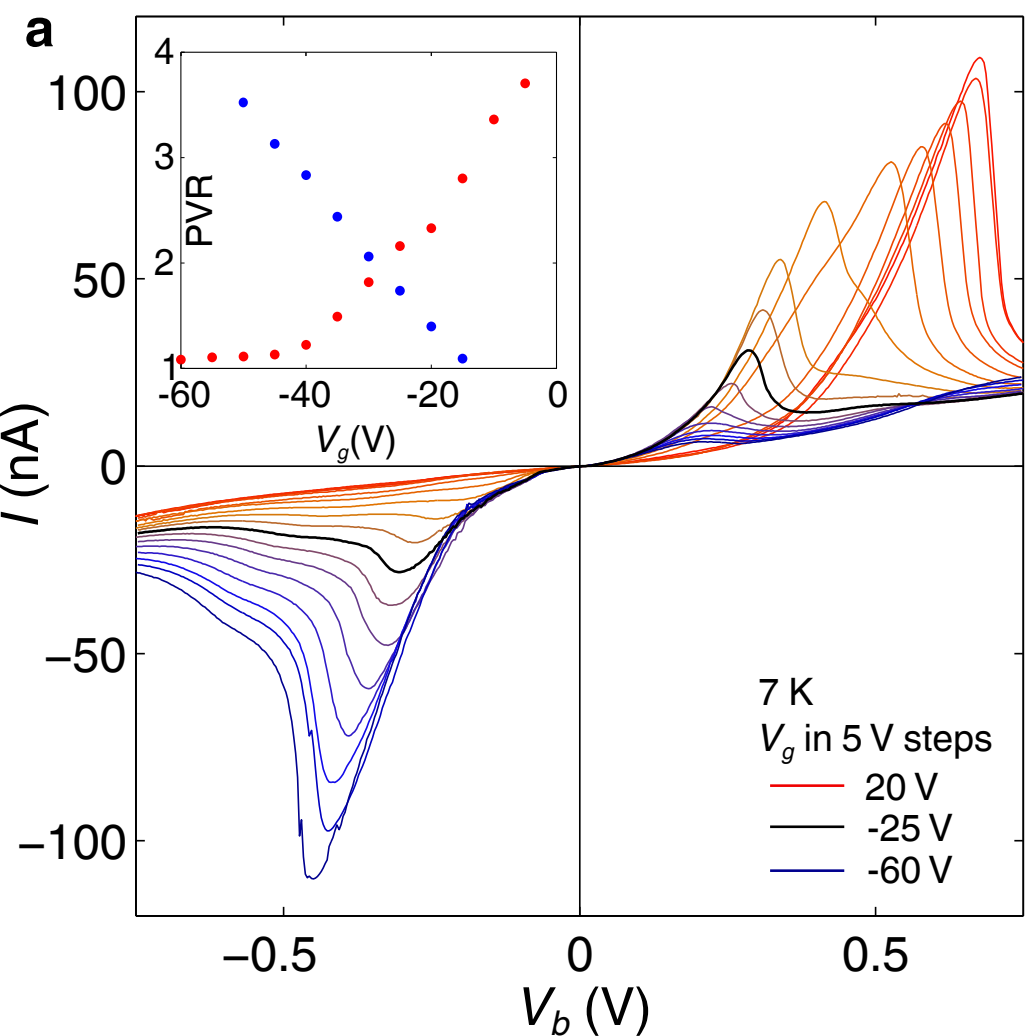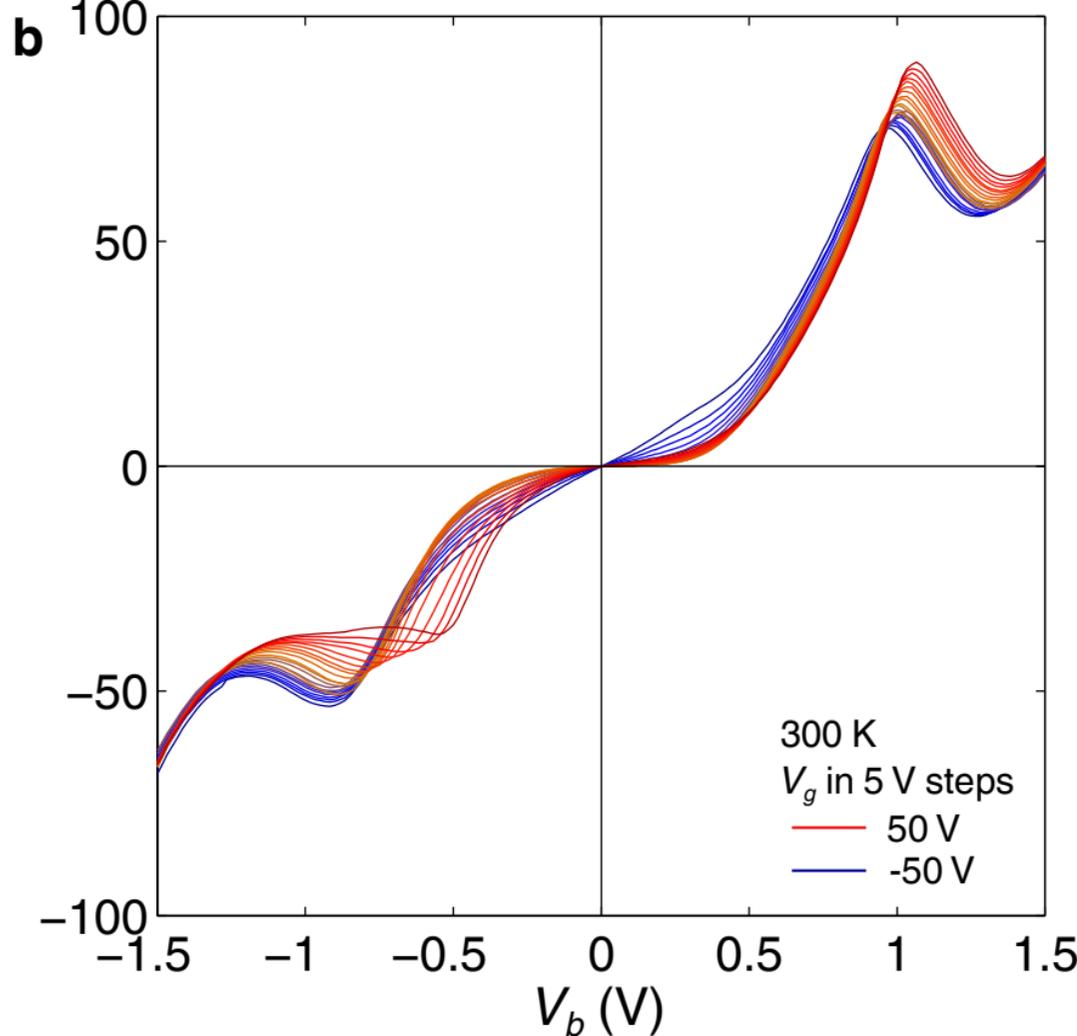